# Interfacial closure of contacting surfaces

F.Rieutord[1,*], C. Rauer[2], H. Moriceau[2]

[1]CEA, INAC, SP2M/NRS, 17 rue des Martyrs, F-38054 GRENOBLE France
[2]CEA, LETI, Minatec Campus, 17 rue des Martyrs, F-38054 GRENOBLE France

**Abstract -** Understanding the contact between solid surfaces is a long standing problem which has a strong impact on the physics of many processes such as adhesion, friction, lubrication and wear. Experimentally, the investigation of solid/solid interfaces remains challenging today, due to the lack of experimental techniques able to provide sub-nanometer scale information on interfaces buried between millimeters of materials. Yet, a strong interest exists improving the modeling of contact mechanics of materials in order to adjust their interface properties (e.g. thermal transport, friction). We show here that the essential features of the residual gap between contacting surfaces can be measured using high energy X-ray synchrotron reflectivity. The presence of this nano-gap is general to the contact of solids. In some special case however, it can be removed when attractive forces take over repulsive contributions, depending on both height and wavelength of asperity distributions (roughness). A criterion for this instability is established in the standard case of van der Waals attractive forces and elastic asperity compression repulsive forces (Hertz model). This collapse instability is confirmed experimentally in the case of silicon direct bonding, using high-energy X-ray synchrotron reflectivity and adhesion energy measurements. The possibility to achieve fully closed interfaces at room temperature opens interesting perspectives to build stronger assemblies with smaller thermal budgets.

**Introduction.** - The bonding strength and energy between assembled materials are dependent on the characteristics of the contacted surfaces. The general situation for solid/solid surfaces (that explains for instance Amontons-Coulomb laws of friction) is that contact occurs only on a small fraction of the apparent surface area, on the top of the highest asperities [1-4]. When clean non-reactive surfaces are considered, attractive van der Waals forces are the only forces left that contribute to the adhesion at room temperature. Van der Waals forces depend strongly on the distance between the attracted bodies, i.e. on the relative morphology of the two surfaces and, at least in a first approximation, on the average distance between the two surfaces [5]. In contacting bodies, the attractive van der Waals forces are balanced by repulsive contact forces that depend on the spatial distribution of the asperities. The balance between these two forces can be made quantitative using models for the van der Waals forces, for the type of contact and for the distribution of asperity heights [6-8]. Such modeling has been performed in the standard case of rough hydrophobic surfaces, transposing standard surface contact mechanics developments at the nanometer scale [9]. At room temperature, solid-solid contact has been described by a balance between elastic non-adhesive repulsive contacts distributed according to a Gaussian statistics and an average van der Waals attraction.

We will show in this letter that the nanometer-wide gap between contacting solid surfaces can be directly measured. These measurements will also show that the standard balance situation leaving an air gap between the solids can be escaped, depending on the type of asperity (densities, frequencies,…), and that situations can be reached where one component (the attractive part here) always dominate over the other, leading to a collapse of the asperity structure and higher bonding energies.

**Silicon direct bonding. -** A textbook example for the study of the contact between surfaces is silicon direct bonding. "Direct bonding" or "direct wafer bonding" is an important technique in material physics as it allows the assembling of a large panel of materials. Contrary to epitaxy, the possibility to realize heterostructures including semiconductors, metals or insulator materials is not dependent on material growth capability (e.g. some crystal lattice matching) but only on the availability of surfaces that are smooth and flat enough. The direct bonding technique has thus many applications, one of the most important industrially being for example the fabrication of silicon-on-insulator (SOI) materials, by direct bonding of single crystalline silicon layer onto an amorphous silicon oxide [10]. Important parameters for the technology are adhesion strength and adhesion energy, as they will condition the possibility or not of subsequent processing of the assembly. Some processes require high bonding energies e.g. to withstand the stress associated to high temperature treatments or handling. Others require on the contrary the possibility to separate back a bonded layer easily (weak strength for debonding). In addition to its technological interest, wafer bonding is also an excellent test vehicle for physics problems of adhesion at nanometer scale [11], both for the statics and dynamics [12].

In the textbook example of hydrophobic silicon (i.e. H-passivated silicon surfaces), the type of distribution for the asperity roughness spectrum can be controlled via the surface preparations. In the following we shall consider two types of hydrophobic silicon surfaces.

1)- "$H_2$-reconstructed" surfaces have been annealed under $H_2$ at high temperature (e.g. 1100°C during 2h) resulting in smooth surfaces exhibiting large-size terraces[13], roughly parallel, whose width is in the 100nm range and height of one atomic step (0.14 nm) (Fig 1a).

2)- "HF-last" surfaces are etched using HF, removing the native oxide and resulting in surfaces whose roughness can be essentially described using Gaussian statistics and whose wavelength spectrum is essentially smooth and centered at high frequencies ($>0.1$ nm$^{-1}$) (Fig.1b)[14]. AFM measurements show that the lateral period of the roughness is short, AFM line profiles showing radii of curvature limited by the tip radius (R<10 nm).

On the experimental side, it is very difficult to study directly the contact interface between standard solids, under load, especially when one wants to go beyond the "integral" adhesion energy value or when the equilibrium distance is to be measured in-situ. The interface then exhibits a nanometer-size narrow gap between asperities, which is embedded between millimeter-thick materials. This gap cannot be measured directly when dealing with the standard case of the contact between flat surfaces between elastic stiff materials. To our knowledge, only artificial patterns involving larger scales using softer materials have been used so far for the direct study of the contact interface between materials. To be able to reach the interface between flat solids, we have extended the standard surface X-ray reflectometry technique to higher energies and to solid/solid interfaces using synchrotron radiation [15]. The technique gives extreme sensitivity to the interface structure, being able to resolve sub-nanometer distance with high accuracy. Typically the technique measures the period and contrast of interference fringes between the two contacting surfaces. The fringe period corresponds to the equilibrium distance of the assembly while the contrast of the fringes relates to the (electron density) filling of the gap. For hydrophobic silicon surfaces, the gap density is very small resulting in a high fringe contrast, and the technique proved to be a sensitive probe to measure the value of gap width and equilibrium distances predicted by the contact physics. We used extensively the technique to characterize the evolution of hydrophilic interfaces upon annealing[16] which allowed e.g. the design of mechanisms for the adsorbed water management[17]. The technique has the advantage of requiring almost no surface preparation, being applicable to actual interfaces, in real-life conditions.

**Asperity contact model.** - The mechanics of contact between solid bodies is at the simplest a balance between van der Waals attractive forces and asperity compression elastic repulsive forces.
The attractive force reads

$$P_{att}(z) = \frac{A}{6\pi z^3} \tag{1}$$

for van der Waals attraction (A is the Hamaker constant for the Si/air/Si).

For the repulsive force, we used here the model of Greenwood and Williamson (GW) [6] where the pressure can be calculated from the relation between the force and the displacement on one individual asperity. This force depends on the geometry and on the elastic properties of the material and generally reads

$$F(\delta) = Ka\,\delta$$

where $a$ is the characteristic width of the contact zone, $\delta$ is the amplitude of the asperity compression and K is a function of elastic parameters (e.g. $K = \frac{4}{3}\frac{E}{1-\nu^2}$ in elastic isotropic materials, where E is the elastic modulus and $\nu$ the Poisson coefficient). Note that the lateral size $a$ of the contact zone depends on $\delta$, $a^2 = \delta R$ in the case of the Hertz model of contact between an elastic sphere of radius $R$ and a rigid plane giving $F(\delta) = K\delta^{3/2}R^{1/2}$.

The repulsive pressure for a distribution of asperities is then obtained summing all the contributions from the asperities at different height levels, hence different levels of compression. It depends on the distribution of summits on each of the facing surfaces $P(z) = \int F(\delta(\zeta,z))n(\zeta)d\zeta$ where $n(\zeta)d\zeta$ is the number of summits per unit area whose height is between $\zeta$ and $\zeta+d\zeta$. In the GW model, n($\zeta$) is supposed to be Gaussian.

The GW model may appear oversimplified as it does not take into account the interactions between neighboring asperities, which play an important role for the standard contact of solid bodies featuring e.g. machined surfaces. Other approaches based on a fractal description of surfaces and including all length scales give an improved description in this case [18-21]. However, the classical GW model offers a fair description of the repulsive part in our case, at least for the HF-last case [9]. A possible explanation is the large lateral wavelength-to-height ratio of our silicon surfaces. Even in the case of the "short" wavelength HF-last surfaces, rms roughness amplitude is in the subnanometer range while the lateral distances between asperities (as can be estimated for example on Fig.1b AFM image) are typically two orders of magnitude larger. Thus one may expect a reduced influence of the stress field interactions between neighboring asperities. Experimentally, the Gaussian statistics for the asperities describes quite well the bearing curve of the HF-last surface obtained from AFM images [9].

The interaction between two rough surfaces can be modeled by the surface/plane interaction provided the roughness σ of the surfaces and the elastic coefficient are renormalized ($\sigma^{*2}=\sigma_1^2+\sigma_2^2$, $1/K^*=1/K_1+1/K_2$). We shall adopt this view in the following and consider plane/surface interactions only to model the interaction between two surfaces whose roughness are uncorrelated.

For a distribution of elastically compressed asperities, the general expression for the repulsive pressure reads

$$P_{rep}(z) = K^* \frac{\sigma^*}{L} F_{3/2}\left(\frac{z}{\sigma^*}\right) \qquad (2)$$

An important parameter we consider in the analysis below is the average distance between asperities L (i.e. the inverse square root of the areal summit density). The exact shape of the function $F_{3/2}(\zeta)$ depends on the height distribution envisioned. At large z, the statistics of heights dominates (as it conditions the number of asperity in contact) and, for the Gaussian statistics considered in the GW model,

$$F_{3/2}(\zeta) = \frac{1}{\sqrt{2\pi}} \int_0^\infty u^{3/2} \exp\left(-\frac{(u+\zeta)^2}{2}\right) du \qquad (3\text{-}1)$$

At small z, nearly all summits are in contact and $F_{3/2}$ approaches unity. The situation is close to that of a regular (periodic) surface where the individual asperity compression regime dominates (Eq.1) and

$$F_{3/2}(\zeta) \approx (1-\zeta)^{3/2} \qquad (3\text{-}2).$$

We have used these expressions to study the balance between this repulsive force and the attractive van der Waals force. This is illustrated Fig. 2.

For small roughness period L (below 10nm typically), the first equilibrium point starting from large separation distances occurs at a large distance compared to roughness ($d_1$). This is a stable equilibrium point. A second equilibrium point could be reached in principle, e.g. forcing a further reduction of distance by exerting an additional external pressure. This point would be unstable and lead to a similar situation as the one described below.

If we now soften the elastic response of the material increasing the roughness wavelength parameter L, the repulsive curve will eventually remain below the attractive part at any z (Fig.2). In this case, the system will compress the asperities till full contact. Let us mention again that the models used for attractive or repulsive forces are strictly not valid at the very short distances corresponding to full contact between solids. They are only used in the analysis below to provide expressions in an essentially dimensional analysis.

The threshold situation can be evaluated using expressions above

$$\begin{cases} P_{att}(z) = P_{rep}(z) \\ \dfrac{dP_{att}(z)}{dz} = \dfrac{dP_{rep}(z)}{dz} \end{cases} \quad (4)$$

Using the power law expressions (1), (2) and (3-2), equations (4) readily give the relation between roughness amplitude and wavelength in the boundary situation, ignoring prefactors:

$$L \approx \frac{K}{A}\sigma^4 \quad (6)$$

In our case, taking $A = 2\ 10^{-19}$ J, $K^* = 10^{11}$ Pa and $\sigma^* = 3\ 10^{-10}$ m gives L in the 10-nm range. This is a criterion for total bondability in the case of van der Waals forces.

Similar relation would be obtained when other expressions are used for asperity compression, e.g. using a plastic compression assumption which would result in a linear dependence of repulsion upon compression in eq. 3.2. In the case of adhesive contacts with local adhesion energy, this balance between adhesion and elastic repulsive forces has been performed and compared to experiments in the case of hydrophilic silicon bonding[22]. Due to the presence of water [15,23,24], the full closure of the interface is never achieved in this case, even though the criterion may explain the absence of macroscopic bonding when one neglects long range van der Waals forces.

We see that if we manage both to have the lowest possible roughness amplitude σ (entering as a power 4, this is a very sensitive parameter) and a large wavelength (i.e. big terraces) we may reach the condition for a complete collapse of the asperities.

**Experimental results and discussion.** - We have performed experiments in regimes corresponding to both situations via surface preparations, "$H_2$-reconstructed" and "HF-last". To study the residual gap, we performed interfacial X-ray reflectivity i.e. reflectivity experiments where the beam enters and exits the sample by the sides [15]. Reflections from external surfaces with large index gradients are thus eliminated (Fig. 3a insert). In addition, electron density profile extraction from reflectivity data can be directly performed by inverse Fourier transform as symmetry of the profile can be assumed. Such technique requires the use of hard energy X-rays so as to limit the absorption of X-rays through sample crossing. In this case, 27 keV (wavelength=0.04592 nm) X-rays from the ESRF BM32 beamline have been used. Samples were cut from standard 300-mm wafer assemblies (total thickness 1.57 mm) in the form of 5 mm-wide samples.

Reflectivity curves from samples prepared using two different surface treatments are shown Fig. 3a. The reflectivities (when plotted in the standard $q^4R(q)$ vs q mode to remove the Fresnel decay with q the wavevector transfer) display large fringes in reciprocal space associated to narrow gaps. The profiles corresponding to these gaps are shown Fig. 3b. They can be described in a fair approximation using a classical rectangular box model, i.e. the interfacial bonding region is assumed to have a constant density. The transition between the different layers is described by a standard rough interface model with a Gaussian statistics, i.e. using an error function profile. The width observed for the two different surface preparations are very different.

In the case of HF-last surfaces, the distance between the two surfaces ($d_1$=0.84 nm) is large compared to the rms roughness of the contacting surfaces (σ=0.2 nm, as measured by AFM or X-ray reflection). This indicates that the mechanical equilibrium occurs by compression of the few highest asperities which are strong enough to resist van der Waals attraction (Fig. 4a). The depth of the gap is close to the silicon density which indicates also a weak overlap of the two roughness systems. Both the equilibrium distance and the density at the gap center are fully consistent with the balance between Si/Air/Si van der Waals forces and an elastic compressive repulsion of a statistical Gauss distribution of asperities with a rms width of 0.2 nm. The measured adhesion energy ($E_1$=19±5 mJ/m$^2$) is also consistent with the work of separation of these two components (i.e. the area between the van der Waals and short-wavelength repulsive force-distance curves of Fig.2). The energy is measured using the standard blade insertion technique [8]. This situation corresponds to the standard contact between rigid materials as e.g. in friction experiments of mechanical parts. In our case, the standard micron range roughness of mechanical surfaces is scaled down to nanometer range, while external load is replaced by van der Waals attraction.

For reconstructed surfaces on the contrary, large terraces are visible whose period is in or above the 100nm range and may be tuned by the control of the vicinal wafer cut angle. The average roughness amplitude given by AFM measurements is slightly smaller than for HF-last samples ($\sigma^*$=0.2 nm). Yet the bonded structure is very different: The width obtained is very small ($d_2$=0.25 nm FWHM). The profile can be described using a Gauss function. Note that due to the large wavevector range of these synchrotron reflectometry measurements, the width measurement accuracy is very high (below 0.01nm). Several measurements performed on different stripes cut in the same wafer samples, or from wafer to wafer show also a good reproducibility (below 0.05nm). The electron density at the gap center is consistent with two overlapping layers of hydrogen at a density of one hydrogen per Si atom on each surface (monohydride species) (Fig. 4b). It should be noted that the electron density profile exhibits small density maxima located next to the gap. This density increase is observable on all samples having received such surface treatments. It should be associated with the surface reconstruction of Si, and its order of magnitude is consistent with the inward motion (.02 nm typically) of the topmost silicon layers[25]. This illustrates the sensitivity of the method. Associated to this reduced distance $d_2$, the adhesion energy $E_2$ is also much increased. Experimentally $E_2$=150±10 mJ/m$^2$, from crack opening measurements, measuring the debonded length in anhydrous atmosphere. This value is again in line with the work of separation that can be calculated using both attractive and repulsive parts of the force: Considering the longer range of the attractive van der Waals part only (which is dominant in the collapsed regime) $A/12\pi d^2$, one obtains a factor of $(d_1/d_2)^2$=10 in the energies, consistent with the observed $E_2/E_1$ ratio. Note that more exact formalisms developed from ab-initio calculations[26] predicted energies in the range of the observed one for the collapsed situation, in contradiction with measurements that were done using HF-last surface samples. Hence, the reported discrepancies between energy calculations and measurements could be manifestations of the effect of short wavelength surface roughness with the presence of an asperity gap.

The possibility to achieve energies in the 0.1J/m$^2$ range without annealing opens also new technological possibilities for the use of direct bonding.

The possibility to achieve full contact also impacts the temperature behavior of the interface. Classically, to improve the interface strength, assemblies are annealed at high temperature. The standard mechanism for gap closure upon annealing is the following. Upon thermal activation, contact points become adhesive (chemical bonds are able to form at contact points) so that interface sealing progresses via a zip-lock mechanism (Fig. 4). This is the standard mechanism for hydrophilic bonding [17]. In our case, a similar mechanism is at work. The experimental signature of such mechanism is a change in the gap density while the gap width remains essentially unchanged. For terraced surfaces, due to the good contact at RT, the reaction can take place at lower temperature, resulting in a full Si/Si bonding at lower temperature (350°C) than for standard rough contact surfaces (700°C). This is directly confirmed experimentally, comparing the interface closure and bonding energy temperature evolutions (Fig. 5).

Finally, it can be pointed out that the transition we describe here for rough solid/solid interfaces has some similarity with the transition from Cassie-Baxter to Wenzel states in the case of wetting of a rough solid by a liquid. This analogy has already been pointed out in the contact of patterned soft adhesive elastomers [27].

**Conclusion. -** We have demonstrated that the standard situation of an asperity-driven partial contact model, generally accepted for most solid/solid contacts, is valid down to nanometer-scale. It can be disputed though when asperities have specific structural features. The criterion can be met in the case of silicon/silicon bonding via appropriate surface reconstruction. In addition to being a textbook example of a contact mechanics effect, the full contact situation has several interesting practical consequences such as an increased adhesion energy at room temperature or a reduced thermal budget to achieve full covalent bonding.

*****

We thank SOITEC for financial support. The support of the European project AtMol is acknowledged (grant agreement N° 270028).

**Figure captions**

**Fig. 1:** Atomic Force Microscopy Images of a) hydrogen annealed reconstructed Si surface 5µmx5µm b) HF-last Si surfaces 1µmx1µm. Color scale is similar, steps on a) have 1.4Å heights.

**Fig. 2:** Attractive (van der Waals) and repulsive (short- and long-wavelength elastic asperity compression) pressure components (Log scale). The $d_1$ distance (0.8nm) is the balance point on the graph between short wavelength asperity repulsion and vdW attractive pressure. The dashed line corresponds to the repulsive component at the transition between a collapsed situation and an asperity-borne repulsion.

**Fig. 3: a.)** reflectivity curves of the two interfaces built by bonding: HF-last prepared surfaces (dots), $H_2$-annealed terraced surfaces (crosses). Thin lines are fits to the data using a simple layer model for the gap between the solids. Insert: geometry of the interface X-ray reflectivity. The X-ray interference is between the amplitude reflected by the two interface gradients. b) electron density profiles (normalized to silicon electron density) as obtained from fig.3a data, showing the two equilibrium distance $d_1$ and $d_2$.

**Fig. 4:** Gap closure mechanism for asperity contacting surfaces. With temperature, the contact becomes adhesive which drives its spreading to the whole interface area.

**Fig. 5**: Annealing temperature evolution of the bonding energy (left scale, circle, solid lines) and the interface gap filling (right scale, triangles, dotted lines). Open symbols are for HF-last surfaces while full symbols are for reconstructed surfaces. The temperatures corresponding to adhesion energy increase due to covalent bonding coincide with gap closure temperatures in both cases.

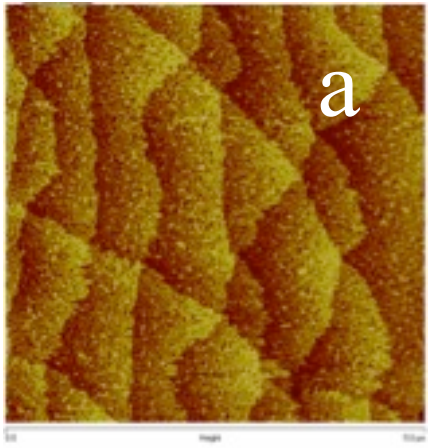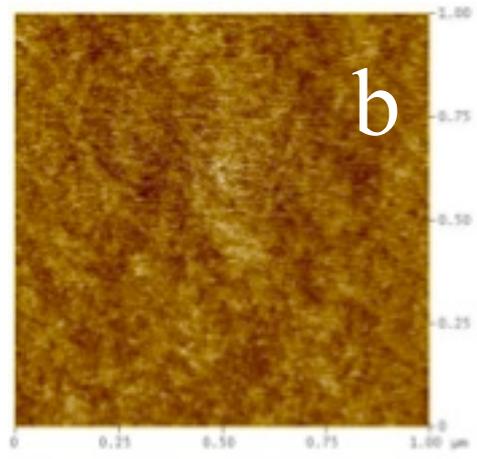

Fig.1

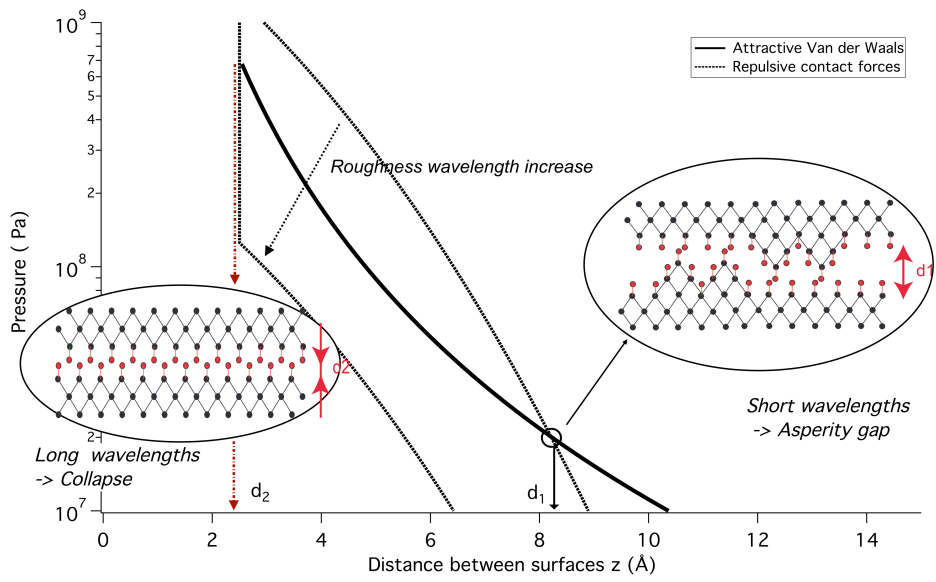

Fig.2

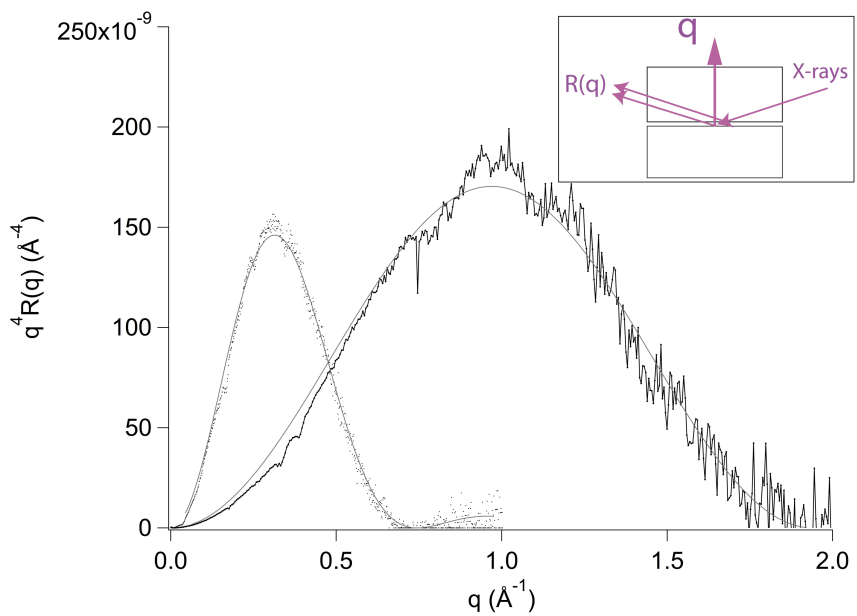

Fig.3a

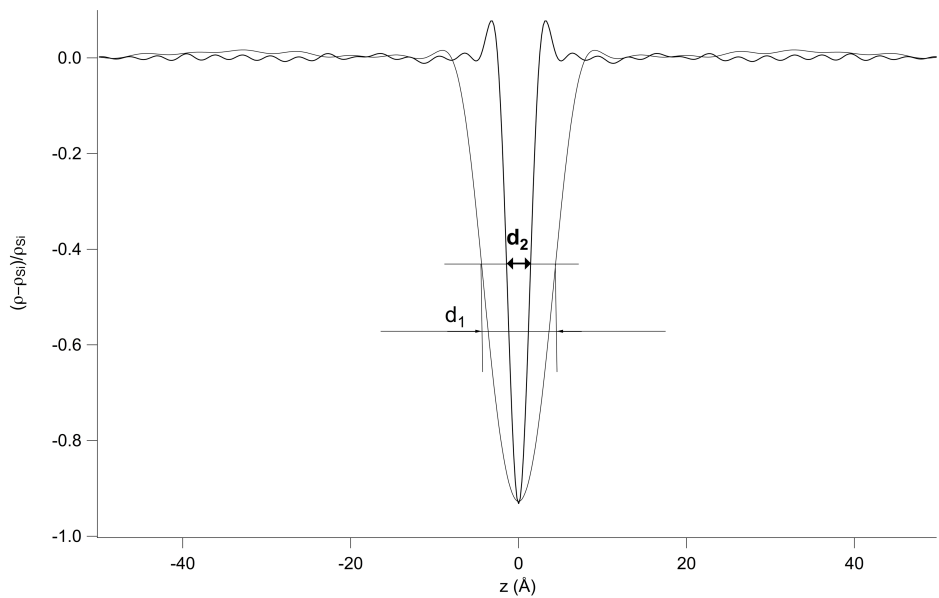

Fig. 3b

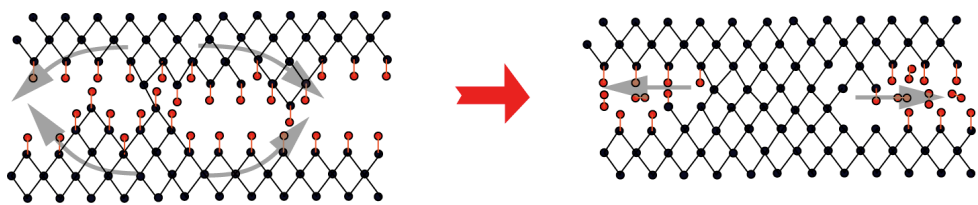

Fig.4

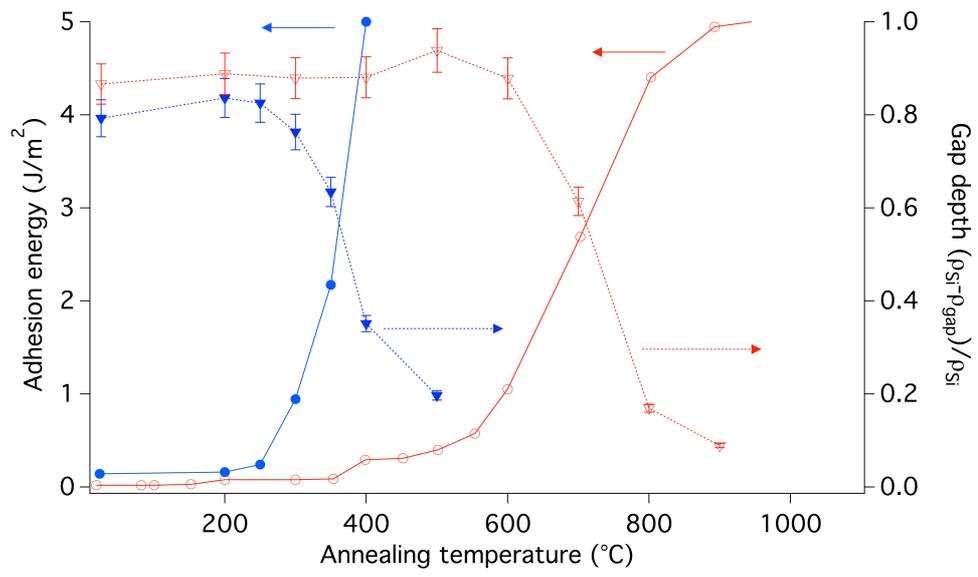

Fig.5